\documentclass[prl,twocolumn,a4paper,superscriptaddress,preprintnumbers,showpacs]{revtex4}
\usepackage{graphicx}
\usepackage{graphicx}
\usepackage{epstopdf}
\usepackage{xspace}
\usepackage{dcolumn}
\usepackage{amssymb,amsmath}
\graphicspath{{eps/}}
\newlength{\figwidth}
\newcommand{\bb}{\ensuremath{B \overline{B}}\xspace}
\newcommand{\btosssss}{\ensuremath{b \to s \overline{s} s
         \overline{s} s} \xspace}
\newcommand{\btophiphik}
     {\ensuremath{B \to \phi \phi K}\xspace}
\newcommand{\bptophiphik}
     {\ensuremath{B^\pm \to \phi \phi K^\pm}\xspace}
\newcommand{\btophiphiks}
     {\ensuremath{B^0 (\overline{B}{}^0) \to \phi \phi K^0_S}\xspace}
\newcommand{\btokphiphi}{\btophiphik}
\newcommand{\bptokphiphi}{\bptophiphik}
\newcommand{\btophiphikp}{\bptophiphik}
\newcommand{\btoksphiphi}{\btophiphiks}

\newcommand{\btoksff}{\btoksphiphi}

\newcommand{\btoetack}{\ensuremath{B^\pm \to \eta_c K^{\pm}}\xspace}

\newcommand{\mb}{\ensuremath{M_{\mathrm{bc}}}\xspace}
\newcommand{\de}{\ensuremath{\Delta E}\xspace}

\newcommand{\lr}{\ensuremath{\mathcal{LR}}\xspace}

\def\mySpecialText{DRAFT $Id$}

\def\myspecial#1{}                   

\myspecial{!userdict begin /bop-hook{gsave 40 150 translate 90 rotate
     /Times-BoldItalic findfont 18 scalefont setfont
     0 0 moveto 0.70 setgray
     (\mySpecialText)
     show grestore}def end}

\begin{document}

\preprint{Belle Preprint 2003-5}
\preprint{KEK Preprint 2003-18}

\title{\boldmath Observation of $B \to \phi \phi K$}

\affiliation{Aomori University, Aomori}
\affiliation{Budker Institute of Nuclear Physics, Novosibirsk}
\affiliation{Chiba University, Chiba}
\affiliation{Chuo University, Tokyo}
\affiliation{University of Cincinnati, Cincinnati, Ohio 45221}
\affiliation{University of Hawaii, Honolulu, Hawaii 96822}
\affiliation{High Energy Accelerator Research Organization (KEK), Tsukuba}
\affiliation{Hiroshima Institute of Technology, Hiroshima}
\affiliation{Institute of High Energy Physics, Chinese Academy of Sciences, Beijing}
\affiliation{Institute of High Energy Physics, Vienna}
\affiliation{Institute for Theoretical and Experimental Physics, Moscow}
\affiliation{J. Stefan Institute, Ljubljana}
\affiliation{Kanagawa University, Yokohama}
\affiliation{Korea University, Seoul}
\affiliation{Kyoto University, Kyoto}
\affiliation{Kyungpook National University, Taegu}
\affiliation{Institut de Physique des Hautes \'Energies, Universit\'e de Lausanne, Lausanne}
\affiliation{University of Ljubljana, Ljubljana}
\affiliation{University of Maribor, Maribor}
\affiliation{University of Melbourne, Victoria}
\affiliation{Nagoya University, Nagoya}
\affiliation{Nara Women's University, Nara}
\affiliation{National Kaohsiung Normal University, Kaohsiung}
\affiliation{National Lien-Ho Institute of Technology, Miao Li}
\affiliation{Department of Physics, National Taiwan University, Taipei}
\affiliation{H. Niewodniczanski Institute of Nuclear Physics, Krakow}
\affiliation{Nihon Dental College, Niigata}
\affiliation{Niigata University, Niigata}
\affiliation{Osaka City University, Osaka}
\affiliation{Osaka University, Osaka}
\affiliation{Panjab University, Chandigarh}
\affiliation{Peking University, Beijing}
\affiliation{Princeton University, Princeton, New Jersey 08545}
\affiliation{University of Science and Technology of China, Hefei}
\affiliation{Seoul National University, Seoul}
\affiliation{Sungkyunkwan University, Suwon}
\affiliation{University of Sydney, Sydney NSW}
\affiliation{Tata Institute of Fundamental Research, Bombay}
\affiliation{Toho University, Funabashi}
\affiliation{Tohoku Gakuin University, Tagajo}
\affiliation{Tohoku University, Sendai}
\affiliation{Department of Physics, University of Tokyo, Tokyo}
\affiliation{Tokyo Institute of Technology, Tokyo}
\affiliation{Tokyo Metropolitan University, Tokyo}
\affiliation{Tokyo University of Agriculture and Technology, Tokyo}
\affiliation{Toyama National College of Maritime Technology, Toyama}
\affiliation{Utkal University, Bhubaneswer}
\affiliation{Virginia Polytechnic Institute and State University, Blacksburg, Virginia 24061}
\affiliation{Yokkaichi University, Yokkaichi}
\affiliation{Yonsei University, Seoul}
  \author{H.-C.~Huang}\affiliation{Department of Physics, National Taiwan University, Taipei} 
  \author{K.~Abe}\affiliation{High Energy Accelerator Research Organization (KEK), Tsukuba} 
  \author{K.~Abe}\affiliation{Tohoku Gakuin University, Tagajo} 
  \author{T.~Abe}\affiliation{Tohoku University, Sendai} 
  \author{I.~Adachi}\affiliation{High Energy Accelerator Research Organization (KEK), Tsukuba} 
  \author{H.~Aihara}\affiliation{Department of Physics, University of Tokyo, Tokyo} 
  \author{M.~Akatsu}\affiliation{Nagoya University, Nagoya} 
  \author{T.~Aso}\affiliation{Toyama National College of Maritime Technology, Toyama} 
  \author{V.~Aulchenko}\affiliation{Budker Institute of Nuclear Physics, Novosibirsk} 
  \author{T.~Aushev}\affiliation{Institute for Theoretical and Experimental Physics, Moscow} 
% ...
\collaboration{Belle Collaboration}

\begin{abstract}
We report the observation of the decay mode $B \to \phi \phi K$ based
on an analysis of 78 fb$^{-1}$ of data collected with the Belle detector
at KEKB.  This is the first example of a \btosssss
transition.  The branching fraction for this decay is
measured to be $\mathcal{B}(B^\pm \to \phi \phi K^\pm) =
(2.6 ^{+1.1}_{-0.9} \pm 0.3) \times 10^{-6}$ for a $\phi\phi$ invariant 
mass below 2.85 GeV/$c^2$.  Results for other related
charmonium decay modes are also reported.
\end{abstract}


\maketitle


We report the observation of the decay mode $B \to \phi \phi K$,
the first example of a \btosssss transition.  In the Standard
Model (SM), this decay channel
requires the creation of an additional final $s\overline{s}$ 
quark pair than in $b \to s \overline{s} s $ processes,
which have been previously observed in modes such as $B \to \phi K$.
In addition to improving our understanding of charmless $B$ decays, 
the $\phi \phi K$ state may be sensitive to glueball production in $B$ 
decays, where the glueball decays to $\phi \phi$ \cite{glueball}.
In addition, with sufficient statistics, 
the decay $B \to \phi \phi K$ could be used to search for a possible  
non-SM $CP$-violating phase in the $b \to s$ transition 
\cite{hazumi}.  Direct $CP$ violation could be enhanced to 
as high as the $40\%$ level if there is
sizable interference between transitions due to non-SM physics
and decays via the $\eta_c$ resonance.
%
%

We use a 78 fb$^{-1}$ data sample 
collected with the Belle detector at the KEKB
asymmetric-energy $e^+ e^-$ (3.5 on 8 GeV) collider \cite{kekb}
operating at the $\Upsilon(4S)$ resonance ($\sqrt{s} = 10.58$ GeV).
The sample contains $85.0 \times 10^6$ produced \bb pairs.
The Belle detector is a large-solid-angle
magnetic spectrometer consisting of a three-layer silicon
vertex detector, a 50-layer central drift chamber (CDC), a
system of aerogel threshold \v{C}erenkov counters (ACC),
time-of-flight scintillation counters (TOF), and an array of
CsI(Tl) crystals located inside a superconducting solenoid
coil that provides a 1.5~T magnetic field.  An iron flux-return
located outside of the coil is instrumented to identify $K^0_L$
and muons.  The detector is described in detail elsewhere
\cite{belle}.


We select well measured charged tracks that have impact parameters with 
respect
to the nominal interaction point (IP) that are less than 0.2 cm in the 
radial direction and less than 2 cm along the beam direction ($z$).
Each track is identified as a kaon or a pion according to a $K/\pi$
likelihood ratio, $\mathcal{L}_K/(\mathcal{L}_\pi + \mathcal{L}_K)$,
where $\mathcal{L}_{K(\pi)}$ are likelihoods derived from
responses of the TOF and ACC systems and $dE/dx$ measurements in the 
CDC.
We select kaon candidates by requiring $\mathcal{L}_K/(\mathcal{L}_\pi
+ \mathcal{L}_K) > 0.6$.
This requirement is 88\% efficient for kaons with a $8.5\%$
misidentification rate for pions.
Kaon candidates that are electron-like according to the
information recorded in the CsI(Tl) calorimeter are rejected.

Candidate $\phi$ mesons are reconstructed via the $\phi \to
K^+ K^-$ decay mode; we
require the $K^+ K^-$ invariant mass to be within $\pm 20$
MeV$/c^2$ ($\pm 4.5$ times the full width)
of the $\phi$ mass \cite{pdg}.
For the \btoksff decay mode,  we use 
$K_S^0\to \pi^+\pi^-$ candidates in the mass window
$482 \,\mathrm{MeV}/c^2 < M(\pi^+\pi^-) < 514 \,\mathrm{MeV}/c^2$
($\pm 4 \sigma$),
where the distance of closest approach between the two
daughter tracks is less than $2.4$ cm,
the magnitude of the impact parameter of each track in the radial
direction exceeds $0.02$ cm, and the flight length
is greater than 0.22 cm. The difference in the
angle between the pion-pair vertex direction from the IP and its 
reconstructed flight direction in the $x-y$ plane is
required to   
be less than 0.03 radians.

To isolate the signal, we form the beam-constrained mass,
  $M_{\rm bc}=\sqrt{E_{\rm beam}^2-|\vec{P}_{\rm recon}|^2}$, and
the energy difference $\Delta E= E_{\rm recon}-E_{\rm beam}$.
Here $E_{\rm beam}$ is the beam energy, and
$E_{\rm recon}$ and $\vec{P}_{\rm recon}$
are the reconstructed energy and momentum of the signal candidate in 
the $\Upsilon(4S)$ center-of-mass frame.
The signal region for $\Delta E$ is $\pm 30$ MeV which corresponds to
$\pm 3.1 \sigma$, where $\sigma$ is the resolution determined from a
Gaussian fit to the Monte Carlo (MC) simulation.
The signal region for \mb is $5.27 \,\mathrm{GeV}/c^2 < \mb < 5.29 
\,\mathrm{GeV}/c^2$. The beam-constrained mass 
resolution is 2.8 MeV/$c^2$, which is mostly due to the
beam energy spread of KEKB.

The major background for the $B\to\phi\phi K$ process is from continuum 
$e^+e^-\to q \bar{q}$ production, where $q$ is a light quark ($u$, $d$, 
$s$, or $c$). Several event topology variables are used
to discriminate the continuum background, which  tends to be collimated 
along the original quark direction, from the more isotropic \bb events.
Five modified Fox-Wolfram moments, the $S_\perp$ variable \cite{sperp} 
and the cosine of the thrust angle are combined into a Fisher
discriminant \cite{SFW}.  
We form signal and background probability density functions (PDFs) for this Fisher discriminant and for the cosine of
the $B$ decay angle with respect to the $z$ axis ($\cos\theta_B$)
for the signal MC and sideband ($5.20 
\,\mathrm{GeV}/{c^{2}} < \mb < 5.26 \,\mathrm{GeV}/{c^{2}}$ and $0.1 < 
|\de| < 0.2$ GeV) data, respectively. 
The  PDFs are multiplied together to form
signal and background likelihoods, ${\cal L}_S$ and ${\cal L}_{BG}$. The
likelihood ratio $\lr \equiv {\cal L}_S/({\cal L}_S+{\cal L}_{BG})$ is
then required to be greater than 0.1.  This requirement retains 97\% of 
the signal while removing 55\% of the continuum background.

\begin{figure}
\includegraphics[clip, trim=0 0 30 40, width=\columnwidth]{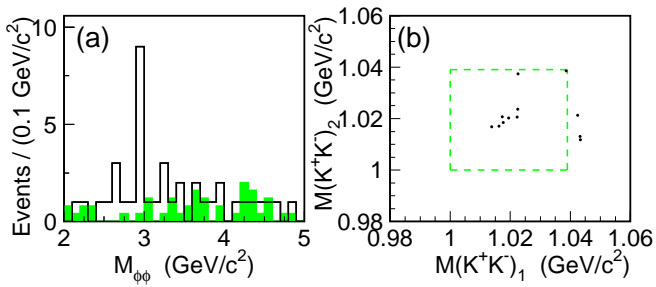}
\caption{\label{fig:mphiphi}
(a) $\phi \phi$ invariant mass spectrum.
The open histogram corresponds to events from the 
\btophiphikp signal region
and the
shaded histogram corresponds to events from the \mb-\de sidebands.
(b) $M_{K^+K^-}$ of one $\phi$ meson candidate versus $M_{K^+K^-}$ of
the other for the events satisfying $M_{\phi\phi} < 2.85$ GeV/$c^2$.  
The
events are concentrated near the point $(M_\phi, M_\phi)$.  The dashed
box shows the region selected for the \btophiphik analysis.
}
\end{figure}

Figure~\ref{fig:mphiphi}(a) shows the $\phi\phi$ invariant mass spectrum 
for events in the \btophiphikp signal region,
where a clear $\eta_c$ peak and some
excess in the lower mass region are evident.

To extract signal yields,
we apply an unbinned, extended maximum likelihood (ML) fit to 
the events with $|\de| < 0.2$ GeV and $\mb > 5.2$
GeV$/c^2$.  The extended likelihood for a sample of $N$ events is
$\mathcal{L} = e^{-(N_S + N_B)} {\prod_{i=1}^N} (N_S \mathcal{P}_i^S +
         N_B \mathcal{P}_i^B )$,
where $\mathcal{P}_i^{S(B)}$ describes the probability for
candidate event $i$ to belong to the signal (background), based on
its measured \mb and \de values.  The exponential factor in the
likelihood accounts for Poisson fluctuations in the total number
of observed events $N$.  The signal yield $N_S$ and the number of 
background events $N_B$ are obtained by maximizing $\mathcal{L}$.
The statistical errors correspond to unit
changes in the quantity $\chi^2 = - 2 \ln \mathcal{L}$ around its
minimum value.  The significance of the signal is defined as the
square root of the change in $\chi^2$ when constraining the number
of signal events to zero in the likelihood fit; it reflects the
probability for the background to fluctuate to the observed event
yield.

The probability $\mathcal{P}$ for a given event $i$ is calculated as
the product of independent PDFs 
for \mb and \de.  The signal PDFs are represented by a Gaussian 
for \mb and a
double Gaussian for \de.  The background PDF for \de is a linear
function; for the \mb background
we use a phase-space-like function with an empirical 
shape \cite{argus}.  The parameters of the PDFs are determined from
high-statistics MC samples for the signal and sideband data for the
background.

\begin{figure}
\includegraphics[width=\columnwidth]{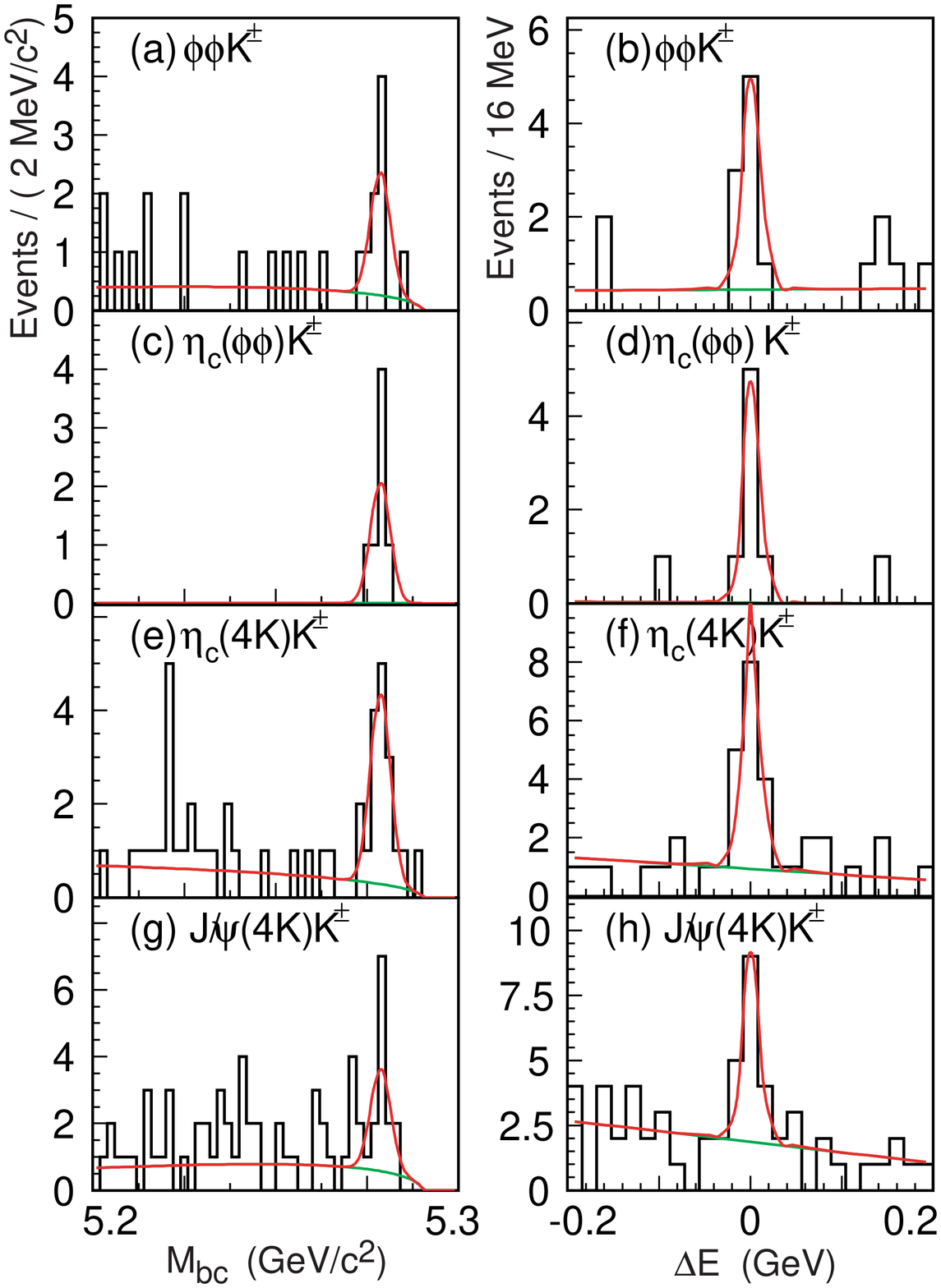}
\caption{\label{fig:2dfit} Projections of \mb and $\Delta E$
overlaid with the fitted curves for (a, b) $\bptokphiphi$ with
$M_{\phi\phi} < 2.85$ GeV$/c^2$, (c, d) \btoetack and $\eta_c
\to \phi \phi$, (e, f) $B^\pm \to \eta_c K^\pm$ and $\eta_c \to 2(K^+
K^-)$, and (g, h) $B^\pm \to J/\psi K^\pm$ and $J/\psi \to 2(K^+
K^-)$.}
\end{figure}

For $M(\phi\phi) < 2.85~\mathrm{GeV}/{c^2}$, the region below the charm
threshold, the ML fit gives an event yield of $7.3 ^{+3.2}_{-2.5}$ with a
significance
of 5.1 standard deviations ($\sigma$).  Projections of the \de
distribution  (with $5.27 \,\mathrm{GeV}/c^2 < \mb < 5.29 
\,\mathrm{GeV}/c^2$)
and of the \mb distribution (with $|\Delta E| < 30$ MeV) are shown in
Figs.~\ref{fig:2dfit}(a,b).  As a consistency check,
a ML fit to the projected \de distribution 
(Fig.~\ref{fig:2dfit}(b) only gives a signal
yield of $7.5 ^{+3.3}_{-2.7}$  with a $4.8 \sigma$ statistical 
significance.
Figure~\ref{fig:mphiphi}(b) shows a scatter plot of the two
$K^+K^-$ invariant masses for events in the $B$ meson 
signal region with $M(K^+K^-K^+K^-) < 2.85~\mathrm{GeV}/{c^2}$
with the $\phi$ mass requirements relaxed.  Here there is a clear
concentration in the overlap region of the two $\phi$ bands. 
There is no event excess in the $\phi$ mass sidebands, which
leads us to conclude that the observed signal is entirely due
to \bptophiphik.  Using a  
signal efficiency of $3.3 \%$, obtained from a
large-statistics MC that uses three-body phase space 
to model the \bptophiphik decays,
we determine the branching fraction for charmless
\bptophiphik with $M_{\phi\phi} < 2.85 \, \mathrm{GeV}/c^2$ to be
\[
\mathcal{B}(\bptokphiphi) = (2.6 ^{+1.1}_{-0.9} \pm 0.3) \times 10^{-6} 
\,,
\]
where the first error is statistical and the second is systematic.


Contributions to the systematic error
include the uncertainties due to the tracking efficiency ($5.4\%$),
particle identification efficiency ($5\%$), and the modeling of
the likelihood ratio cut ($2\%$).  The error due to the modeling
of the likelihood ratio cut is determined using $B^- \to D{}^0 (\to
K^- \pi^+ \pi^- \pi^+) \pi^-$ events in the same data sample; these
events have the same number of final-state particles and an 
event topology that is similar to the \bptokphiphi signal.
The uncertainty due to the MC $M_{\phi\phi}$ modeling ($4\%$)
accounts for the  $M_{\phi\phi}$ dependence of the
detection efficiency.  The systematic error in the
signal yield ($6\%$) is determined by varying the means and
$\sigma$ of the signal and the shape parameters of the background.
We determine an upper limit of 5\% on the
possible contamination by non-resonant 
$B^\pm \to \phi (K^+ K^-)_{\mathrm{NR}} K^\pm$ or 
$B^\pm \to 2 (K^+ K^-)_{\mathrm{NR}} K^\pm$ decays by 
redoing the fits with the  $\phi$ mass requirement relaxed.
The sources of systematic error are combined in quadrature to
obtain the final systematic error of $12\%$.

For the \btoksphiphi mode, there are only four signal candidates.
We combine the \bptokphiphi and \btoksphiphi modes and perform a
ML fit and obtain a signal event 
yield of $8.7 ^{+3.6}_{-2.9}$ with $5.3 \sigma$ statistical
significance.
Assuming isospin symmetry, we obtain
\[
\mathcal{B}(\btokphiphi) = (2.3 ^{+0.9}_{-0.8} \pm 0.3) \times 10^{-6} 
\,,
\]
for $M_{\phi\phi} < 2.85 \, \mathrm{GeV}/c^2$.


No enhancement is observed in the $M_{\phi\phi}$ region 
corresponding to the $f_J(2220)$ glueball candidate~\cite{pdg}, 
also refered to as $\xi$.   Assuming the mass and width of $f_J(2220)$ to 
be 2230 MeV/$c^2$ and 20 MeV/$c^2$,  we define a signal region of 
$2.19 \,\mathrm{GeV}/c^2 < M_{\phi\phi} < 2.27 \,\mathrm{GeV}/c^2$,
$5.27 \,\mathrm{GeV}/c^2 < \mb < 5.29 \,\mathrm{GeV}/c^2$ and $|\de| < 
30$ MeV.  One event is observed in this region with an expected 
background, estimated from the sideband, of 0.5.   Using an extended 
Cousins-Highland method that uses the the Feldman-Cousins 
ordering scheme and takes systematic uncertainties into 
account~\cite{pole},  we 
obtain a $90\%$ confidence level (CL) upper limit of 3.7 signal events,
which corresponds to
\[
         \mathcal{B}(B^\pm \to f_J(2220) K^\pm) \times
         {\mathcal{B}}(f_J(2220) \to \phi \phi) <
         1.2 \times 10^{-6}.
\]


We select \btoetack, $\eta_c \to \phi \phi$ candidates
by requiring
$2.94 \,\mathrm{GeV}/c^2 < M_{\phi\phi} < 3.02 \,\mathrm{GeV}/c^2$.
A clear signal is
evident in Figures~\ref{fig:2dfit}(c,d), and the fitted yield of
$N_S = 7.0 ^{+3.0}_{-2.3}$ events has a significance of $8.8 \sigma$.  
The corresponding branching fraction is
\begin{eqnarray*}
\lefteqn{\mathcal{B}(B^\pm \to \eta_c K^\pm) \times
     \mathcal{B}(\eta_c \to \phi \phi)} \hspace{1in} \\
     & & = (2.2 ^{+1.0}_{-0.7} \pm 0.5) \times 10^{-6}.
\end{eqnarray*}
In addition to the previously listed sources of systematic error,
here the error also includes the possible contamination from
charmless \bptophiphik decays, which is estimated to be less than 1.2
events.
Using the measured branching fraction $\mathcal{B}(B^\pm \to \eta_c
K^\pm) = (1.25 \pm 0.42) \times 10^{-3}$~\cite{etack}, we determine the 
$\eta_c \to \phi \phi$ branching fraction to be
\[
\mathcal{B}(\eta_c \to \phi \phi)
     = (1.8 ^{+0.8}_{-0.6} \pm 0.7) \times 10^{-3} \,,
\]
which is smaller than the current world average value of $(7.1 \pm 2.8) 
\times
10^{-3}$~\cite{pdg}.

\begin{table*}
\begin{ruledtabular}
\caption{\label{bf} Signal yields, efficiencies including secondary
branching fractions, statistical
significances and branching fractions of \btophiphik and related
decays.  The branching fractions for modes with $K^+ K^-$ pairs include
contributions from $\phi \to K^+ K^-$.}
\begin{tabular}{lcddc}
Mode    & Yield
         & \multicolumn{1}{l}{Efficiency (\%)}
     & \multicolumn{1}{c}{Significance ($\sigma$)}
     & $\mathcal{B}$ ($\times 10^{-6}$)  \\ \hline
$B^\pm \to \phi \phi K^{\pm}$ ($M_{\phi\phi} < 2.85$ GeV$/c^2$)
     & $7.3 ~^{+3.2}_{-2.5}$
     & 3.3 & 5.1
     & $ 2.6 ~^{+1.1}_{-0.9} \pm 0.3$ \\
$B \to \phi \phi K$ ($M_{\phi\phi} < 2.85$ GeV$/c^2$)
     & $8.7 ~^{+3.6}_{-2.9}$
     & 2.2 & 5.3
     & $ 2.3 ~^{+0.9}_{-0.8} \pm 0.3$ \\
$B^\pm \to f_J(2220) K^\pm$, $f_J(2220) \to \phi\phi$
     & $< 3.7$
     & 3.6
     & .
     & $< 1.2$ \\
$B^\pm \to \eta_c K^\pm$, $\eta_c \to \phi \phi$
     & $7.0 ~^{+3.0}_{-2.3}$
     & 3.7 & 8.8
     & $2.2 ~^{+1.0}_{-0.7} \pm 0.5$  \\
$B^\pm \to \eta_c K^\pm$, $\eta_c \to \phi K^+ K^-$
     & $14.1 ~^{+4.4}_{-3.7}$
     & 4.6 & 7.7
     & $3.6 ~^{+1.1}_{-0.9} \pm 0.8$  \\
$B^\pm \to \eta_c K^\pm$, $\eta_c \to 2(K^+ K^-)$
     & $14.6 ~^{+4.6}_{-3.9}$
     & 9.6 & 6.6
     & $1.8 ~^{+0.6}_{-0.5} \pm 0.4$  \\
$B^\pm \to J/\psi K^\pm$, $J/\psi \to \phi K^+ K^-$
     & $9.0 ~^{+3.7}_{-3.0}$
     & 4.4 & 5.3
     & $2.4 ~^{+1.0}_{-0.8} \pm 0.3$	\\
$B^\pm \to J/\psi K^\pm$, $J/\psi \to 2(K^+ K^-)$
     & $11.0 ~^{+4.3}_{-3.5}$
     & 9.2 & 4.8
     & $1.4 ~^{+0.6}_{-0.4} \pm 0.2$  \\
\end{tabular}
\end{ruledtabular}
\end{table*}

\begin{figure}
\includegraphics[width=\columnwidth]{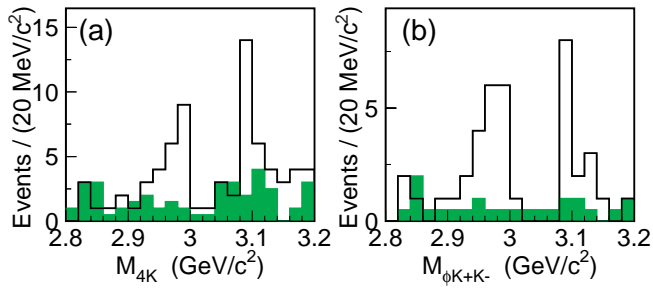}
\caption{\label{fig:5k}
(a) $2(K^+ K^-)$ and (b) $\phi K^+ K^-$ invariant mass spectra in
the $\eta_c$ and $J/\psi$ regions.  The open histograms correspond to
events from the
$B$ signal region, and the
shaded histograms correspond to events from the \mb-\de sidebands.
}
\end{figure}

Since the $J/\psi$ and $\eta_c$ charmonium resonances also decay to
$2(K^+ K^-)$, the decay chains $B \to \mathrm{charmonium} +
K$ with charmonium $\to 2(K^+ K^-)$ can provide consistency checks 
of the $B \to \phi \phi K$ analysis.
To select $B \to 2 (K^+ K^-) K$ candidates, we apply tighter particle
identification and continuum suppression requirements than in
the case of \btophiphik in order to reduce the larger combinatoric 
background.  Figure~\ref{fig:5k}(a) shows the invariant mass distribution 
of any two pairs of $K^+ K^-$, $M_{4K}$, between 2.8 GeV$/c^2$ and 3.2 
GeV$/c^2$ for the events in the $B$  signal region.
Significant contributions from both $\eta_c$ and $J/\psi$ intermediate
states are seen.

To identify the signals from $\eta_c$ and $J/\psi$ intermediate states,
we require that the invariant mass of $2(K^+ K^-)$ satisfy
$2.94 \,\mathrm{GeV}/c^2 < M_{4K} < 3.02 \,\mathrm{GeV}/c^2$
and $3.06 \,\mathrm{GeV}/c^2 < M_{4K} < 3.14 \,\mathrm{GeV}/c^2$,
respectively.  We use signal yields from  ML fits to determine 
branching fractions.  Figures~\ref{fig:2dfit}(e--h) show the \mb 
and \de projection plots with the fitted curves superimposed.
Table~\ref{bf} summarizes the signal yields, efficiencies,
statistical significances, and the branching-fraction products.
By requiring the invariant mass of one of the $K^+ K^-$ pairs to
correspond to a $\phi$ meson, we also measure the decays of $B^\pm \to
\eta_c (J/\psi) K^\pm$ and $\eta_c (J/\psi) \to \phi K^+ K^-$.  The
results are included in Table~\ref{bf}.

Using the known branching fractions 
$\mathcal{B}(B^\pm \to J/\psi K^\pm) = (1.01 \pm 0.05) \times 10^{-3}$ 
\cite{pdg} and $\mathcal{B}(B^\pm \to \eta_c K^\pm)$, we obtain the 
secondary branching fractions for $J/\psi$ and 
$\eta_c$ decays to $2(K^+ K^-)$ and $\phi K^+ K^-$ listed in 
Table~\ref{ccbarbf}.

Our measured branching fractions for $\eta_c \to \phi \phi$ and $\eta_c
\to 2 (K^+ K^-)$ are smaller than those of previous experiments \cite{pdg},
while those for $J/\psi$ decays are consistent.
The decay $\eta_c \to 2 (K^+ K^-)$ proceeds dominantly through
$\eta_c \to \phi K^+ K^-$ with $\phi \to K^+ K^-$.  This is the first 
measurement of $\eta_c \to \phi K^+ K^-$.  The decay of $\eta_c \to \phi \phi$ 
with $\phi \to K^+ K^-$ makes up approximately $1/3$ of the branching fraction 
of $\eta_c \to \phi K^+ K^-$. 

\begin{table}
\caption{\label{ccbarbf} Measured branching fractions of secondary
charmonium decays and the world averages \cite{pdg}.   The
branching fractions for modes with $K^+ K^-$ pairs include contributions
from $\phi \to K^+ K^-$.}
\begin{ruledtabular}
\begin{tabular}{lcc}
Decay mode
     & $\mathcal{B}$ (this work)
     & $\mathcal{B}$ (PDG)   \\ \hline
$\eta_c \to \phi\phi$
     & $(1.8 ~^{+0.8}_{-0.6} \pm 0.7) \times 10^{-3}$
     & $(7.1 \pm 2.8) \times 10^{-3}$ \\
$\eta_c \to \phi K^+ K^-$
     & $(2.9 ~^{+0.9}_{-0.8} \pm 1.1) \times 10^{-3}$
     & -- \\
$\eta_c \to 2(K^+ K^-)$
     & $(1.4 ~^{+0.5}_{-0.4} \pm 0.6) \times 10^{-3}$
     & $(2.1 \pm 1.2) \;\%$  \\
$J/\psi \to \phi K^+ K^-$
     & $(2.4 ~^{+1.0}_{-0.8} \pm 0.3) \times 10^{-3}$
     & $(7.4 \pm 1.1) \times 10^{-4}$  \\
$J/\psi \to 2(K^+ K^-)$
     & $(1.4 ~^{+0.5}_{-0.4} \pm 0.2) \times 10^{-3}$
     & $(7.0 \pm 3.0) \times 10^{-4}$
\end{tabular}
\end{ruledtabular}
\end{table}


In summary, we have observed the charmless three-body decay
\btophiphik, which is the first example of a \btosssss transition.
The branching fraction 
$\mathcal{B}(\bptokphiphi)
= (2.6 ^{+1.1}_{-0.9} \pm 0.3) \times 10^{-6}$ 
for $M_{\phi\phi} < 2.85 \,\mathrm{GeV}/c^2$,
is measured with significances of $5.1 \,\sigma$.
No signal is observed for the decay $B \to f_J(2220) K$ with $f_J(2220) 
\to \phi \phi$.  The corresponding upper limit at 90\% C.L. is 
$\mathcal{B}(B^\pm \to f_J(2220) K^\pm) \times {\mathcal{B}}(f_J(2220) 
\to \phi \phi) < 1.2 \times 10^{-6}$.
We have also observed significant signals for $B^\pm \to \eta_c K^\pm$ 
with
$\eta_c \to \phi\phi$, with $\eta_c \to \phi K^+ K^-$, and with
$\eta_c \to 2(K^+ K^-)$, as well as a signal for $B^\pm \to J/\psi 
K^\pm$ with $J/\psi \to
\phi K^+ K^-$.  We report the first measurement of $\eta_c
\to \phi K^+ K^-$ with a branching fraction of
$\mathcal{B}(\eta_c \to \phi K^+ K^-)
     = (2.9 ^{+0.9}_{-0.8} \pm 1.1) \times 10^{-3}$.
Our measured branching fractions for $\eta_c \to \phi \phi$ and
$2(K^+K^-)$ are smaller than those of previous experiments.

\begin{acknowledgments}
We wish to thank the KEKB accelerator group for the excellent
operation of the KEKB accelerator.
We acknowledge support from the Ministry of Education,
Culture, Sports, Science, and Technology of Japan
and the Japan Society for the Promotion of Science;
the Australian Research Council
and the Australian Department of Industry, Science and Resources;
the National Science Foundation of China under contract No.~10175071;
the Department of Science and Technology of India;
the BK21 program of the Ministry of Education of Korea
and the CHEP SRC program of the Korea Science and Engineering 
Foundation;
the Polish State Committee for Scientific Research
under contract No.~2P03B 01324;
the Ministry of Science and Technology of the Russian Federation;
the Ministry of Education, Science and Sport of the Republic of 
Slovenia;
the National Science Council and the Ministry of Education of Taiwan;
and the U.S.\ Department of Energy.
\end{acknowledgments}


\bibliography{phiphik-prl}

\end{document}